\documentclass[conference]{IEEEtran}
\IEEEoverridecommandlockouts
\usepackage{booktabs}
\usepackage{cite}
\usepackage{amsmath,amssymb,amsfonts}
\usepackage{algorithmic}
\usepackage{graphicx}
\usepackage{textcomp}
\usepackage{listings}
\usepackage{xcolor}
\usepackage{verbatim}
\usepackage{subcaption}
\usepackage{caption}
\usepackage[normalem]{ulem}
\usepackage[ruled,vlined]{algorithm2e}
\newcommand{\todo}{\textcolor{red}}

\def\BibTeX{{\rm B\kern-.05em{\sc i\kern-.025em b}\kern-.08em
    T\kern-.1667em\lower.7ex\hbox{E}\kern-.125emX}}
\SetKwBlock{KwState}{state:}{}
\SetKwProg{Proc}{procedure}{}{} 
\begin{document}
\title{ADaPT: \underline{A}daptive-window \underline{D}ecoding for \underline{P}ractical fault-\underline{T}olerance\\
% \thanks{Correspondance: toberoi@uchicago.edu}
}
% \author{\normalsize{QCE 2026 Submission}}

\author{
\IEEEauthorblockN{Tina Oberoi\IEEEauthorrefmark{1}, 
                  Joshua Viszlai\IEEEauthorrefmark{1}, 
                  Frederic T. Chong\IEEEauthorrefmark{1}}
\IEEEauthorblockA{\IEEEauthorrefmark{1}University of Chicago, Chicago, Illinois, USA}
\thanks{$^*$Correspondence: \texttt{toberoi@uchicago.edu}}
}

\maketitle
\thispagestyle{plain}
\pagestyle{plain}

\begin{abstract}
Window decoding, first proposed to reduce decoding complexity for real-time decoding, is an essential component to realize scalable, universal-fault tolerant computation. Prior work has focused on improving throughput through parallelization and reducing reaction time via speculation on window boundaries. However, these methods use a fixed window size $ d$, paying a fixed decoding time overhead for each window. In practice, we find this overhead of a fixed window size unnecessary in many cases due to the sparsity of average-case errors in QEC. Leveraging this insight, in this paper we propose an adaptive window decoding technique based on decoder confidence. This technique reduces the overhead in decoding time thus reducing reaction time without compromising on logical error rates. We benchmark adaptive window decoding across different codes and hardware inspired noise models. Our results show that this adaptive technique reaches the target error rate while maintaining a low decoding time overhead across different codes, and under different noise models. 
\end{abstract}

\begin{comment}
\begin{IEEEkeywords}
quantum error correction, decoding
\end{IEEEkeywords}
\end{comment}

\section{Introduction}
Many problems capable of demonstrating quantum advantage require fault tolerance, necessary to build resilience against hardware noise and errors inherent in current quantum devices. Quantum error correction (QEC) offers this resilience by distributing each logical qubit across many physical qubits and repeatedly performing parity checks. The decoding of these checks is a critical component of QEC responsible for estimating the most likely logical correction. Although decoding is performed on classical hardware, it has significant implications on the performance and scalability of fault-tolerant quantum computations. For example, applying a T gate using teleportation (Fig 1 in \cite{skoric2023parallel}) requires a measurement-based classical correction after teleportation. However, this correction can only be applied reliably once the logical information has been successfully decoded from the encoded state. To achieve accurate inference of logical errors, QEC typically requires $  O(d)  $ rounds of syndrome measurements for reliable decoding. In such cases where reliable and real-time logical decoding is required, window decoding addresses this challenge.
\begin{figure}
    \centering
    \includegraphics[width=\linewidth]{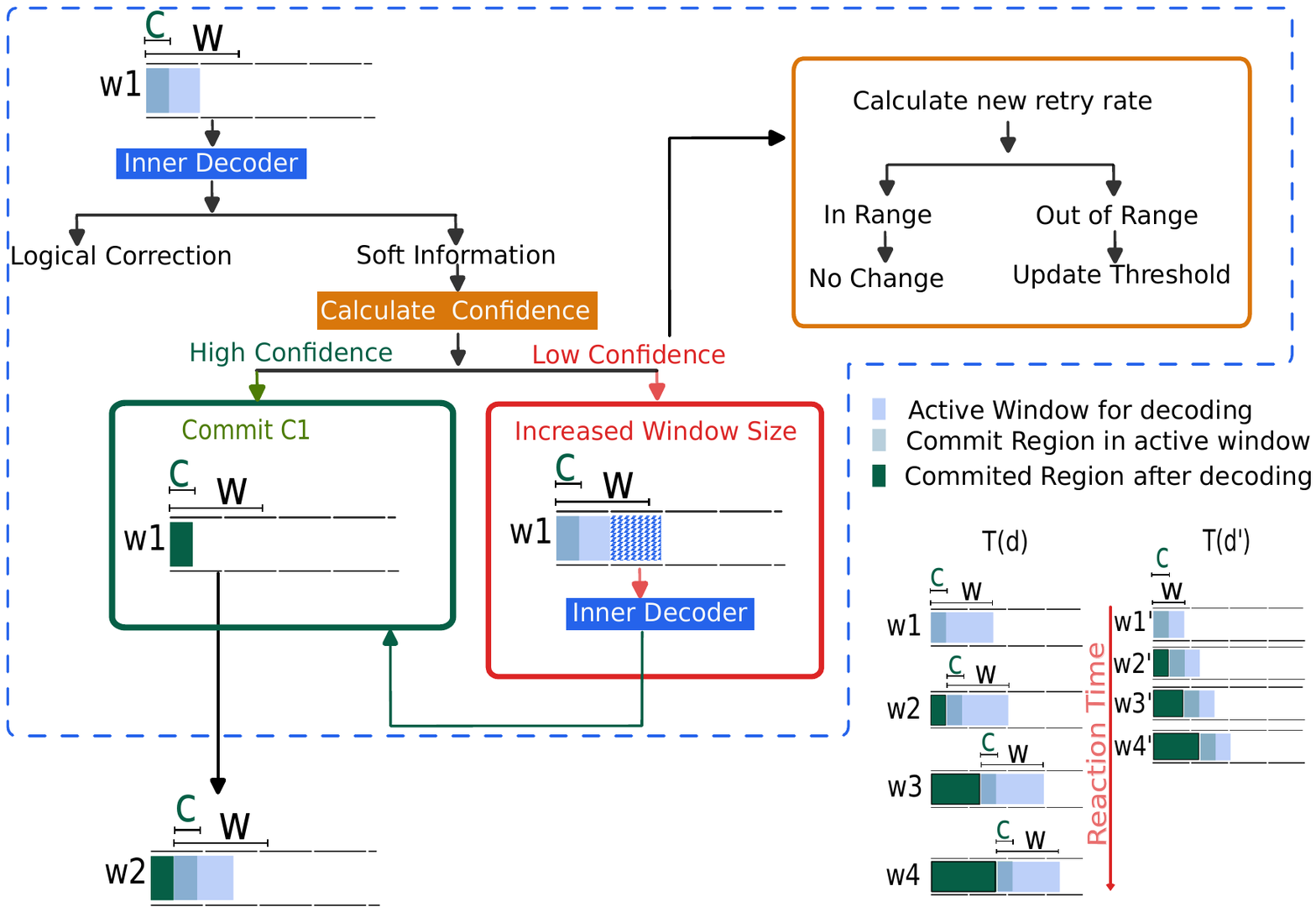}
    \caption{\textbf{An overview of Adaptive Window Decoding} Decoding starts with a small window $  d'  $ to obtain logical correction and soft information. A confidence score is computed; if confidence is high we commit correction $  C  $ and proceed to the next window, otherwise the window is enlarged to $  d  $ and retried. Concurrently, the observed retry rate $  r_{obs}  $ is tracked to dynamically adjust $  Q_{threshold}  $. (Bottom right) \textbf{Decoding with a reduced window size ($  d' < d  $).} The reaction time with $d'$ compared to $d$ follows $  T(d') < T(d)  $, thus reducing window size reduces reaction time.}
    \label{fig:hero_fig}
\end{figure}
The core idea of window decoding is to divide the full syndrome history into smaller overlapping segments, known as windows. Each window is then processed independently by an inner decoder. Although significant progress has been made in improving throughput and reducing reaction time for window decoders, prior approaches have primarily focused on exploiting parallelism with a fixed window size $  W = d  $.

\begin{figure}
    \centering
    \includegraphics[width=0.9\linewidth]{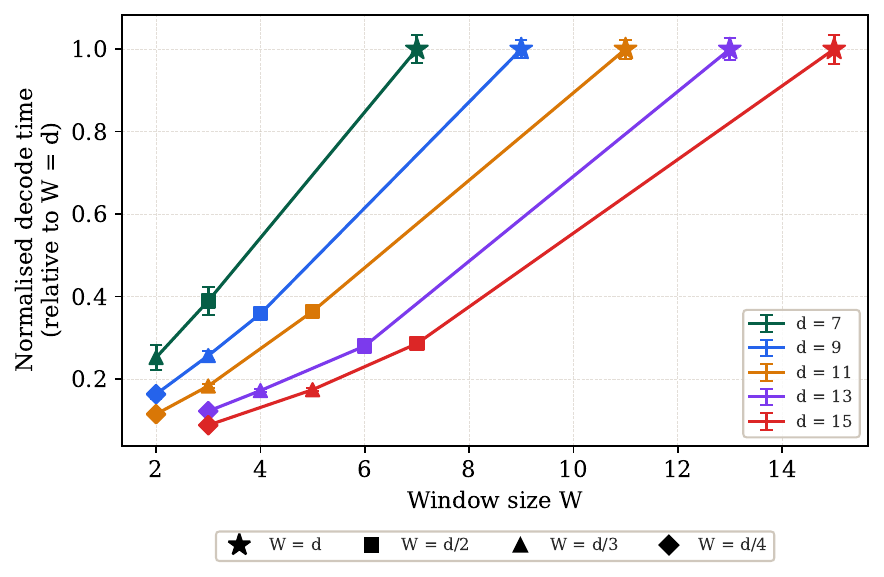}
    \caption{\textbf{Decoding time decreases superlinearly with smaller window sizes.} The decoding time for toric codes with distance $  d=7  $ using the BP+LSD decoder under depolarizing noise. The y-axis shows the decoding time for a window of size $  W  $, normalized by decoding time when $  W = d  $. The x-axis shows the window size $  W  $. Decoding time decreases superlinearly as $  W  $ is reduced.}
    \label{fig:decoding_time}
\end{figure}

A fixed window size forces an unnecessary upper limit on decoding time per window. When errors are sparse, we pay more overhead than required. As the code distance $  d  $ increases, the window size grows proportionally, which in turn increases the complexity for the inner decoder. Consequently, the decoding time for a fixed window size scales super linearly with $  d  $. This behavior can be seen in Fig.~\ref{fig:decoding_time} for the BP+LSD decoder \cite{hillmann2025localized}, a state-of-the-art decoder for QLDPC codes.
% For the BP+LSD decoder, the decoding time scales superlinearly with window size. 
At $  d=7  $, using $W = \lfloor d/2 \rfloor$ yields approximately $2.5\times$ speedup over $  W = d  $. At $  d=15  $, the speedup grows to $3.5\times$.
\begin{comment}
This superlinear scaling is expected, as the computational complexity of BP+LSD increases with both the number of detectors and the number of faults in the window — both of which scale with$  W $ (in worst-case). 
\end{comment}
In this paper, we propose an adaptive window size technique for real-time decoding. This adaptive technique uses a small window size and adaptively increases the window size $  W  $ only when required. In the average-case where a small window size was sufficient, this approach improves the reaction time for window decoding.

The primary tradeoff of a reduced window size is an increased logical error rate (LER). The adaptive method leverages decoder confidence to retry windows with low decoding confidence and re-decode using an increased window size. With a fixed retry budget, we are able to keep a low decoding time overhead while closing the LER gap. We implement this technique on a sliding window decoder, which forms the basis of all improvements in window decoding. Thus, our approach is easily extensible to parallel and speculative window decoders. Our benchmarks on Toric codes and Bivariate Bicycle Codes show that this adaptive technique closes the gap between a baseline (smaller window size) and a target (larger window size) LER, while keeping the decoding time 0.4-0.6x of upper limit ($W=d$). Furthermore, the confidence-based metric we use is derived from the decoder’s soft information, making it applicable to a wide range of quantum error-correcting codes. 

The remainder of this paper is organized as follows. Section~\ref{sec:prereq} provides the necessary background on real-time decoding, reviews prior enhancements in window decoding, and discusses the overhead associated with fixed-size window decoding. Section~\ref{soln} introduces adaptive window decoding algorithm and describes the confidence metric used to dynamically adjust the window size. Section~\ref{sec:results} evaluates the performance of the adaptive window sizing approach across various quantum error-correcting codes and under different noise models. Finally, Section~\ref{sec:conclusion} summarizes our key findings, highlights the importance of low-reaction time in decoding, and outlines promising directions for future research.
\section{Background}\label{sec:prereq}

\subsection{Decoding}
In QEC codes, a logical qubit is encoded into many physical qubits. These physical qubits are divided into data qubits, which carry the logical information, and syndrome (ancilla) qubits, which are repeatedly measured to detect errors on data qubits. The resulting continuous stream of syndrome measurements is fed into a decoder. By processing this syndrome information and inferring appropriate correction operations, the decoder keeps track of and mitigates logical errors that corrupt the logical information. 

To perform this inference, the decoder is provided with the parity-check matrix $  H \in \mathbb{F}_2^{|D| \times |F|}  $ and syndrome \(\mathbf{s}\). Each row of $  H  $ corresponds to a detector $  D  $ (a specific combination of measurement outcomes), and each column corresponds to a possible fault $  F  $. The entry $  H_{df} = 1  $ if fault $  f  $ triggers detector $  d  $ ($\forall d \in D$), and $  0  $ otherwise. The $  H $ matrix establishes a mapping from faults in the quantum circuit to the triggered detectors. A syndrome \(\mathbf{s}\) is a binary vector of length \(|D|\), where each entry corresponds to the outcome of a detector: \(s_i = 0\) indicates no error detected at the \(i\)-th detector, and \(s_i = 1\) indicates an error (for \(i = 1, \dots, |D|\)). This decoding problem can be modeled into a decoding graph, which is a bipartite graph with detectors as nodes and potential errors are represented as edges. Now the problem is reduced to finding the minimum weight error such that $\textbf{s} = H. \hat{\textbf{e}} $. This error vector $\hat{\textbf{e}}$ determines the recovery operator used for logical corrections.
Some of the commonly used decoders are minimum-weight perfect matching (MWPM) \cite{higgott2022pymatching, delfosse2021almost, fowler2012surface} and union-find (UF) \cite{delfosse2021almost}. MWPM assigns edge weights \( w = \log((1-p)/p) \) (where \( p \) is the error probability) in the decoding graph and solves for the minimum-weight matching; and UF combines Pauli and erasure errors into modified erasure $\epsilon$ and resolves them by growing and merging valid clusters. MWPM and UF work well for surface codes. However, they are not applicable to quantum low-density parity-check (QLDPC) codes, which have emerged as a popular choice in QEC due to their superior encoding rate~\cite{bravyi2024high, stein2024architectures}. 

QLDPC codes rely on belief propagation (BP) and its variants for decoding. Standalone BP is often ineffective due to the presence of degenerate errors in quantum codes. BP+OSD (Ordered Statistics Decoding)~\cite{panteleev2021degenerate} and BP+LSD (Localized Statistics Decoding)~\cite{hillmann2025localized} incorporate post-processing when BP does not converge. Both methods leverage inversion decoding, using the beliefs from BP to construct the inversion matrix. Specifically, OSD computes the solution by inverting a full-rank submatrix: $  \hat{\mathbf{e}}_{[I]} = \hat{H}_{[I]}^{-1}. \mathbf{s}  $. In contrast, LSD performs parallel inversions on small cluster submatrices: $  \hat{\mathbf{e}}_{[C_1 \cup C_2 \cup \dots \cup C_k]} = \left( \hat{H}_{[C_1]}^{-1}. \mathbf{s}_1, \, \hat{H}_{[C_2]}^{-1}. \mathbf{s}_2, \, \dots, \, \hat{H}_{[C_k]}^{-1} .\mathbf{s}_k \right)  $, where the clusters are disjoint subsets derived from the graph. These decoders are commonly used as inner decoders in window-based decoding schemes.

\subsection{Soft Information in Decoding}
Besides logical corrections decoder also provides useful soft information. Soft information refers to all additional data provided by the decoder beyond the logical correction itself. This includes error probabilities, likelihoods of different error mechanisms, and cluster information that contribute to the final decoding decision.  Soft information has been used previously to improve decoding in surface codes by initializing weights in matching decoders using the log likelihood ratios (llr) from running a belief propagation instance first~\cite{Higgott_2023}. It is also used in concatenated codes where information from the inner codes after decoding can then be used for outer code decoding \cite{fowler2013optimalcomplexitycorrectioncorrelated, gidney2025yoked}. Soft information can also be used to calculate decoder confidence as a post-selection criteria, for example: the complementary gap used in magic state cultivation circuits~\cite{gidney2024magicstatecultivationgrowing, sahay2026foldtransversalsurfacecodecultivation}. 
Real-time decisions based on decoder confidence can also reduce the overhead of lattice surgery operations~\cite{akahoshi2025runtimereductionlatticesurgery} and improve overall decoding performance by switching between a slow and fast decoder~\cite{toshio2025decoder}.

\subsection{Real Time Decoding}

\begin{comment}
    While providing the decoder with the entire syndrome history enables accurate logical inference for Clifford circuits, it becomes computationally expensive for real-time, reliable logical correction with long syndrome histories.
\end{comment}

In QEC, there are two main decoding paradigms: offline decoding and online (real-time) decoding. Offline decoding processes the entire syndrome history at once to produce a logical correction. Although it can achieve high accuracy by using $  O(d)  $ rounds of syndrome measurements \cite{fowler2012surface}, the computational complexity of handling the full history makes it unsuitable for returning real-time logical corrections.
To enable real-time decoding, sliding window decoding \cite{dennis2002topological, huang2021between, iyengar2012windowed} divides the continuous syndrome stream into smaller, overlapping segments called windows. These windows are processed sequentially by an inner decoder, eliminating the need to feed the entire syndrome history at once. The inner decoder can be MWPM or Union-Find (UF), or BP+OSD / BP+LSD depending on the quantum error-correcting code used.
Each window is defined by three parameters: the window size $  W  $, the commit size (stride) $  C  $, and a buffer region of size $  W - C  $, with the constraint $  W > C  $.
In standard sliding window decoding, the window size is conventionally set to $  W = d  $ for a code of distance $  d  $. This choice ensures performance by properly accounting for measurement errors, while limiting the decoding problem to a manageable size of $  W  $ at each step. After decoding a window, the corrections in the commit region $  C  $ are finalized. Error chains crossing the boundary between the committed and buffer regions are converted into artificial defects and carried forward. The next window is then formed by sliding forward, incorporating these artificial defects along with new syndrome information from subsequent rounds. 

Parallel window decoding \cite{skoric2023parallel} addresses the backlog problem \cite{terhal2015quantum, tan2023scalable, 10.1145/3695053.3731022} — defined as when the rate of syndrome data is generated faster than it can be processed:
$$\frac{\text{rate at which syndrome data is generated}}{\text{rate at which syndrome data is processed}} > 1$$
by splitting the sequential windows into two overlapping streams, $  A  $ and $  B  $. Windows in stream $  A  $ can be decoded concurrently, after which boundary information is passed to stream $  B  $. While this approach significantly improves throughput, the reaction time remains suboptimal because stream $  B  $ must still wait for stream $  A  $ to complete. Reaction time is defined as
$$T_{\text{time at which window was created}} - T_{\text{time at which window was decoded}}$$
Speculative decoding \cite{10.1145/3695053.3731022} mitigates this limitation by enabling early exchange of boundary information, allowing stream $  B  $ to begin decoding almost immediately and thereby reducing reaction time.

\begin{figure}[htbp]
    % \centering
    \includegraphics[height=0.5\textheight, keepaspectratio]{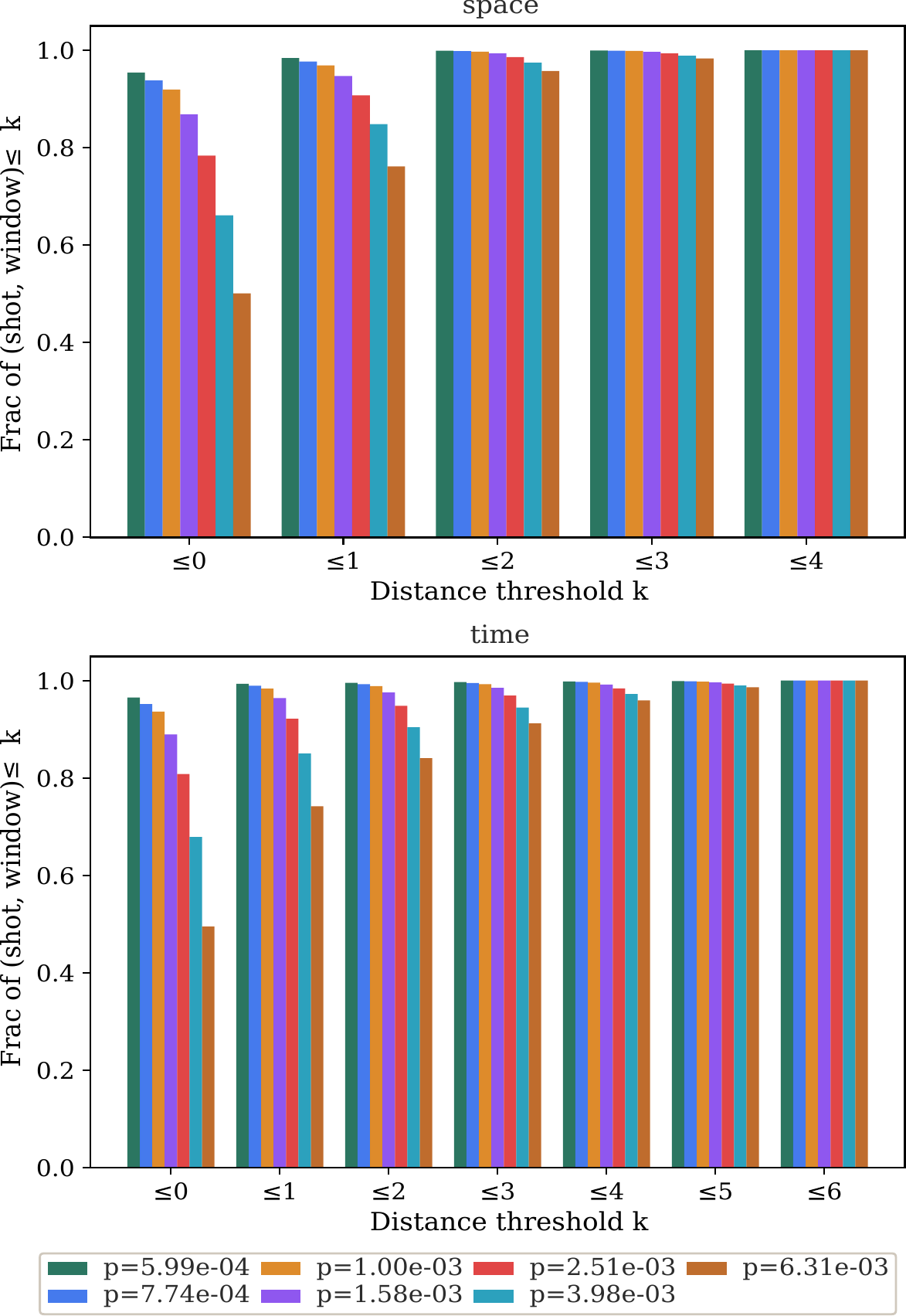}  
    \caption{\textbf{Detector separation statistics in fixed window decoding of the $d=7$ toric code.} Window decoding with $W=d$ is performed on the toric code under depolarizing noise for different $p$. Over all shots and windows, we record the maximum nearest-neighbour distance among triggered detectors — measured in space (2D toric Manhattan, top) and in time    ($|\Delta t|$, bottom). In $\approx90\%$ of cases this maximum is $\leq d/2$, yet a fixed-$W$ decoder incurs the  full overhead of $W=d$ in every window.}
\label{fig:your_label}
    \label{fig:detector_distance}
\end{figure}

\subsection{Window Size Performance}
We analyzed the decoding time for Toric codes for different distance and different window sizes. We used BP+LSD for decoding, and recorded time to decode each window. As evident in Fig~\ref{fig:decoding_time}, the increase in decoding time for increasing window size is superlinear. Despite the above discussed advances, prior works on window decoding relies on a fixed window size $  W = d  $. This fixed-size approach imposes an unnecessary overhead, particularly at low physical error rates where errors are sparse. 
To verify the sparsity argument, we analyze the distribution of inter-detector distances within each window ($W=d$) across shots, for the Toric code ($d=7$). For each (shot, window) pair, we identify all fired X-type detectors and compute each detector's nearest-neighbour (NN) distance to the closest other fired detector in the same window — in space (2D toric Manhattan) or in time ($|\Delta t|$) — and record the maximum. Fig.~\ref{fig:detector_distance} shows the cumulative distribution of this maximum NN distance across all (shot, window) pairs. In both space and time, $\approx 90\%$ of the detector pairs have a maximum nearest-neighbor distance of $\leq 3$ (i.e., $\leq d/2$). 
This is consistent with the sparsity assumption and shows that most windows remain decodable with a reduced window size $W = \lfloor d/2 \rfloor$. 
In low-error (sparse) regions, this reduced window size significantly lowers the reaction time, achieving an approximately $2.5\times$ reduction per round for $d=7$ (see Fig.~\ref{fig:decoding_time}).
The problem with reduced window size is a high LER in cases where the assumption that errors are sparse no longer holds. To resolve this we propose adaptive window sizing for window decoding. 
In this work, we use soft information from the decoder to assess its confidence and dynamically adjust the window size in our adaptive window decoding scheme. 
One common metric for evaluating decoder confidence is the logical gap (also known as the complementary gap) \cite{bombin2024fault, gidney2025yoked}. The complementary gap is computed by augmenting the error graph and toggling the boundary detectors to find the two best matchings — one that matches the boundary and one that does not. The logical gap is then defined as the log-ratio of the weights of these two matchings. A small gap (close to zero) indicates low decoder confidence, while a large gap indicates high confidence. 
But logical gap comes with its own problems \cite{lee2026efficientpostselectiongeneralquantum}:
\begin{enumerate}
    \item Logical gap requires fixing of a logical class which works for MWPM decoder and guarantees minweight correction in this fixed class, which is not the case for other decoders like Union-find. The performance of decoders like union-find significantly degrades if a logical class is fixed and also doesnt guarantee minimum weight.
    \item high computational cost which grows exponentially i.e. $2^{k}$ for $k$ logical observables. This cost becomes intractable for codes encoding multiple logical qubits.
\end{enumerate} 
And alternative for calculating decoder confidence was proposed in \cite{lee2026efficientpostselectiongeneralquantum}. Here authors used decoder's soft information to discard low confidence decoding shots. We will use the same metric to define the confidence of the decoder, and dynamically increase the size of the window when confidence is low.

\begin{comment}
Although this reduces the complexity for the decoder from entire history to windows, this technique suffers from backlog problem \cite{terhal2015quantum}defined as:
$$\frac{\text{rate at which syndrome data is generated}}{\text{rate at which syndrome data is processed}} > 1$$
If unaddressed, the backlog can lead to an exponential slowdown in the execution speed of fault-tolerant quantum computation .
\end{comment}

\begin{comment}
    To address this we use cluster metrics to quantify decoder's confidence for a given problem. By leveraging this approach, we can dynamically expand the window size exclusively for low-confidence decoding instances. While there is no free lunch, the associated retry overhead is incurred only for those specific retried windows. 
\end{comment}

\section{Proposed Solution}
\label{soln}
We start by introducing the two main components of the adaptive window decoding technique. The first is the confidence metric, which forms the basis for dynamically increasing the window size according to the reliability of the model’s predictions. The second is the dynamic hypertuner, which is responsible for keeping the overall decoding time overhead in check. 

\begin{algorithm}[t]
\caption{Dynamic Hypertuner}
\label{alg:hypertuner}

\KwIn{Initial threshold $c_0$, step size $\delta$, target band $[r_{\min}, r_{\max}]$}

$c \leftarrow c_0$\;
$n_{\text{proc}} \leftarrow 0$\;
$n_{\text{retry}} \leftarrow 0$\;

\BlankLine

\SetKwProg{Proc}{procedure}{}{}
\Proc{\textsc{Update}(\textit{retried})}{
    $n_{\text{proc}} \gets n_{\text{proc}} + 1$\;
    \If{\textit{retried}}{$n_{\text{retry}} \gets n_{\text{retry}} + 1$\;}
    
    $r_{\text{obs}} \leftarrow n_{\text{retry}} / n_{\text{proc}}$\;
    
    \If{$r_{\text{obs}} > r_{\max}$}{
        $c \leftarrow c \cdot (1 + \delta)$\;
    }
    \ElseIf{$r_{\text{obs}} < r_{\min}$}{
        $c \leftarrow \max(c \cdot (1 - \delta), 0)$\;
    }
    
    \KwRet $c$\;
}
\end{algorithm}
\begin{figure}[htbp]
    \centering
    \includegraphics[width=0.5\textwidth]{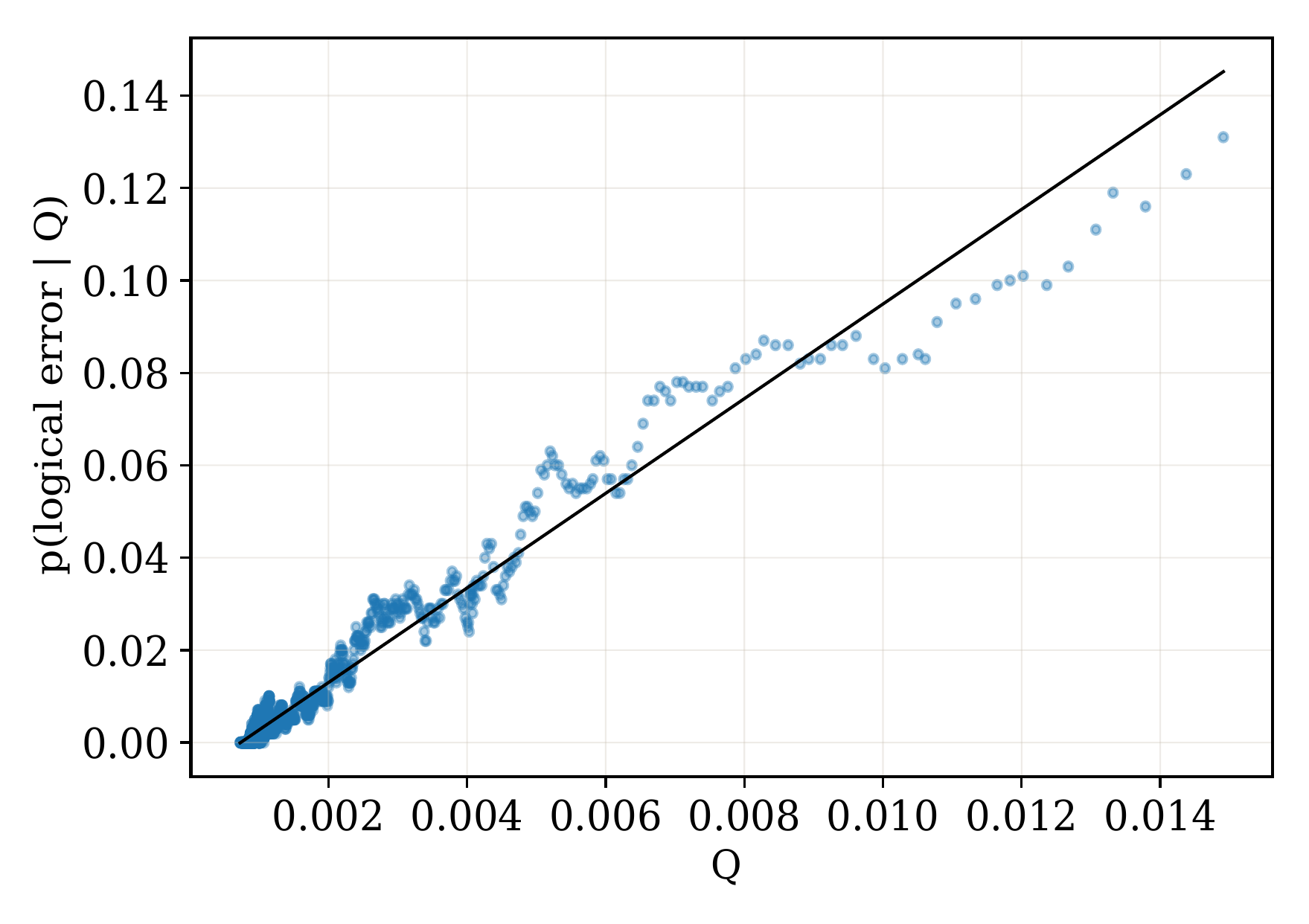}
    \caption{\textbf{Correlation of logical error rate (LER) with cluster-based confidence metric (Q).} 
    This is a memory experiment on the toric code under depolarizing noise with physical error rate $p = 0.005$, 
    decoded using the BP-LSD. Here $Q$ is calculated globally over the whole experiment. 
    The plot shows that the logical error probability increases as $Q_{\text{llr}}$ increases, 
    demonstrating the inverse dependence of LER on $Q$.}
    \label{fig:ler_vs_q}
\end{figure}

\subsection{Cluster Based Confidence Metric}
The alternative to logical gap proposed for quantifying decoding confidence is by using cluster statistics from decoder. The intuition behind this metric is: \textit{small physical errors indicate localized errors, and as the size/llrs of clusters increase the probability of logical failure increases.} Based on this intuition the confidence metric we use here is defined as $\alpha$-norm of error clusters $C_i$ in error mechanisms $\mathcal{E}$:
\begin{equation}
    Q({C_i}; \mathcal{E}) = \frac{1}{\sum_{e \in \mathcal{E}} w_e}(\sum_i( \sum_{e \in C_i} w_e)^\alpha)^{\frac{1}{\alpha}}
\end{equation}
Here, $  w_e  $ denotes the log-likelihood ratio (LLR) associated with error $  e  $. The parameter $  Q  $ is inversely correlated with the decoding confidence — that is, a lower value of $  Q  $ indicates higher confidence in the decoded result. Inspired by stability experiments \cite{Gidney_2022, akahoshi2025runtimereductionlatticesurgery}, Fig~\ref{fig:ler_vs_q} analyzes how the logical error rate of the Toric code varies with the parameter $  Q  $ when decoded using the BP+LSD decoder. Since $Q$ is inversely correlated with decoding confidence, as $Q$ increase the probability of logical error increases. In \cite{lee2026efficientpostselectiongeneralquantum} \( Q \) metric is employed as a post-selection strategy in global and real-time decoding. If any window in sliding window satisfy \( Q > c \) (low confidence), that shot is discarded. We use  \( Q\) to decide confidence of window and retry with increased window size for \( Q > c \) (low confidence). 

This metric allows us to determine whether a smaller window suffices for a decoding problem, balancing time constraints and LER budgets. In Sec~\ref{sec:results} we demonstrate that this strategy achieves a reduced LER with only a modest overhead in decoding time. For our purposes we combine the error clusters from current and committed error clusters from previous window to define confidence. The authors \cite{lee2026efficientpostselectiongeneralquantum} used cutoff as $c = 0.01$ for benchmarks as this value represents the crossover point between low Logical Error Rate (LER) and low discard rate. However, it was observed that this cutoff is highly sensitive to both the code distance $  d  $ (which determines the code’s resilience to physical errors) and the physical error rate itself. Therefore, instead of relying on a static cutoff for every experiment, we introduce a dynamic hyperparameter tuning system to adaptively select an appropriate cutoff value.

\begin{figure}
    \centering
    \includegraphics[width=\linewidth]{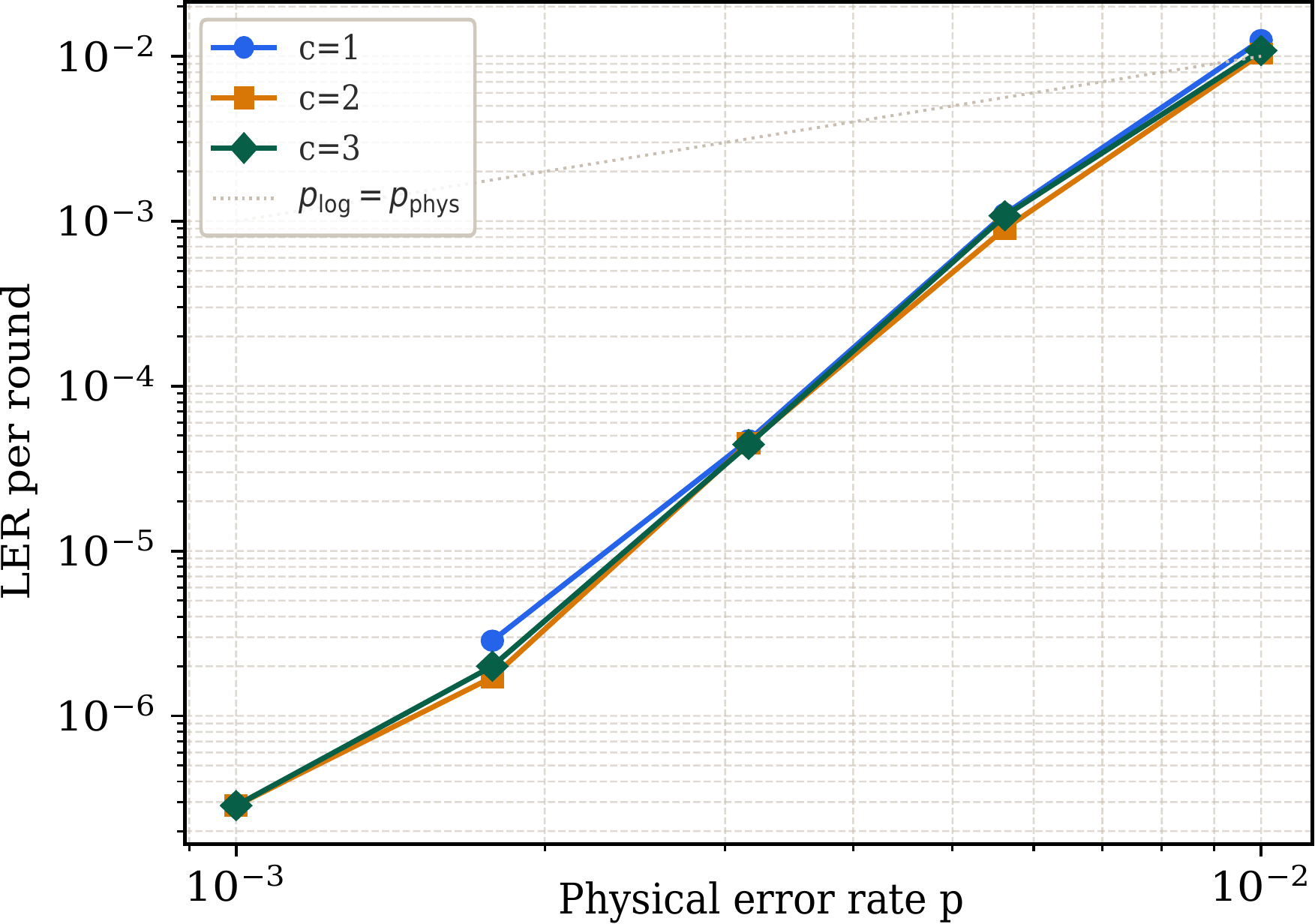}
    \caption{\textbf{Effect of logical error rate on changing commit size}
This figure shows that the logical error rate (LER) remains essentially unchanged for commit sizes $  c \in \{1, 2, 3\}  $ when the buffer size is fixed at $  d-1  $, for sliding-window decoding of the toric code at distance $  d = 7  $.}
    \label{fig:commit_size_comp}
\end{figure}

\subsection{Dynamic Tuning Confidence Cuttoff}
The goal of adaptive window decoding is to balance the tradeoff between LER — pushing it closer to the target window decoder — and timing overhead, which grows with retry frequency. A first approach could be to perform a similar static analysis as \cite{lee2026efficientpostselectiongeneralquantum}, where we do a cutoff sweep for a range ($0.001 - 0.01$), and select cutoff that gives the lowest retry rate and ler. However, a static threshold lacks per-window adaptability and can vary significantly across error rates, code families, window sizes, and noise models. To address this, we adopt dynamic hypertuning inspired by feedback control loops. In this approach, the retry rate, $r$, serves as the controlled variable, with a target band \([r_{\min}, r_{\max}]\) as the setpoint to constrain decoding time overhead. The confidence threshold \(c\) functions as the actuator.

While classical PID controllers incorporate proportional (responding to current error), integral (addressing accumulated steady-state error), and derivative (anticipating future error via rate of change) terms, here we use a simpler on-off control law, focusing solely on proportional adjustments. As outlined in Algorithm~\ref{alg:hypertuner}, it combines a dead zone with a on-off feedback loop. Here, $  r_{\text{obs}}  $ is the observed retry rate, given by the sliding average of retried windows over the total number of windows processed. If (\(r_{\text{obs}}\)) falls within \([r_{\min}, r_{\max}]\), the tuner enters the dead zone and leaves \(c\) unchanged. If the rate deviates outside this band, \(c\) is updated via a fixed multiplicative step:
\[
c \leftarrow c \cdot (1 \pm \delta),
\]
where \(\delta\) is a small constant step size. This design is straightforward and robust: the dead zone prevents oscillations (hunting) near the target, while the fixed step size limits the aggressiveness of threshold adjustments per window.

The observed rate \(r_{\text{obs}}\) is calculated as a global cumulative average across all processed windows. Although a sliding-window average would react more swiftly to recent behavior, we favor the cumulative approach for its natural damping effect — it preserves a stable initial threshold during short runs (e.g., tens of windows) and avoids overreactions to transient fluctuations. We initialize \(c = c_0\) and allow the hypertuner to adapt it online as decoding progresses. Notably, the hypertuner relies solely on cluster statistics already generated by the decoder, eliminating the need for extra sampling or offline sweeps. It also imposes no assumptions about the code, distance, noise model, or error rate, ensuring broad applicability across the settings explored in this work.

\subsection{Adaptive Window Decoding}
In previous sections we saw that a fixed window size of $  W = d  $ offers strong error correction but incurs unnecessary overhead under typical sparse errors. To balance decoding latency and LER, we propose adaptive window decoding as shown in Fig~\ref{fig:hero_fig}. This method begins with a small window and uses cluster information from the inner decoder as a confidence metric ($Q$). If the decoding confidence is low ($ Q > c $) it increases $  W  $ size and re-decodes with the increased window size. Else if the decoding confidence is high, the window is committed and the decoder moves to the next window. Running in parallel the dynamic hypertuner keeps track of the retry rate $r_{obs}$. If the retry rate is within the budget, the threshold remains unchanged. Else the threshold is adjusted proportionally. With a limited retry budget, this approach closes the LER gap between $  W = d/2  $ and $  W = d  $, while reducing average decoding latency to $0.4-0.6\times$  that of the full $  W = d  $.

\begin{comment}
    As shown in Fig\ref{fig:hypertuner_noisemodels}, the results are for Toric code $(d=,p=)$ for different noise models, we can see the reduce windows's LER is reduced by the hypertuner but the overhead still remains much smaller. 
This shows that our tenchique agnostic to the underlying code or hardware constraints can come closer to reducing the decoding time per measurement round. Which was only proven once before for distributed UF decoder on sepcialised FPGA hardware \todo{add namitha's helios paper}

\end{comment}

\section{Results}\label{sec:results}
\begin{figure}
    \centering
    \includegraphics[width=\linewidth]{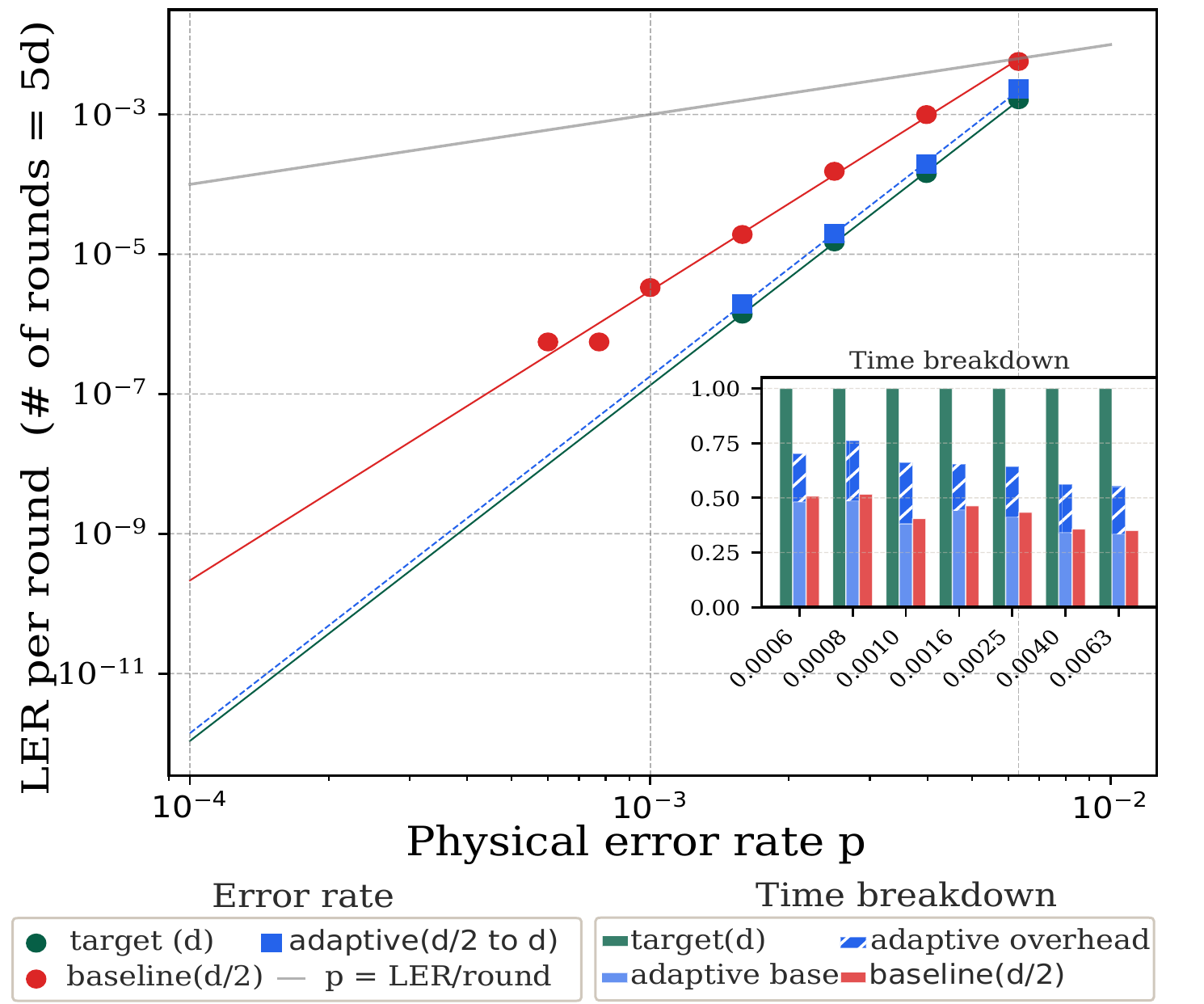}
    \caption{\textbf{Adaptive Window Decoding on $d=7$ Toric code } This Figure illustrates LER per round for the \(d=7\) toric code under depolarizing noise. The physical error rate \(p < 0.01 \) , with 35 rounds of syndrome measurements (\(\approx 5d\)). The baseline employs a small window size ($\lfloor d/2 \rfloor$), the target uses a full window size (\(d\)), and the adaptive approach defaults to the baseline window size ($\lfloor d/2 \rfloor$) but retries only low-confidence windows with the increased size (\(d\)). (Inset) decoding times normalized relative to the target window size $d$ (maximum time).}
    \label{fig:toric_code_ler_vs_p_45}
\end{figure}

Here, we benchmark adaptive window decoding on Toric codes ($d=7$) and QLDPC codes (Bivariate Bicycle Codes \cite{bravyi2024high}, [[72, 12, 6]]).
\begin{comment}
    in the Pauli-Z basis.
\end{comment} 

To conduct simulations, we use the QLDPC repository \cite{perlin2023qldpc} for code generation and  sliding window decoding. We use the BP+LSD decoder \cite{Roffe_LDPC_Python_tools_2022} for inner decoding with cluster statistics provided by the decoder \textit{stats}. For BP+LSD we use $LSD\_0$ and \textit{min\_sum} bp method. Circuit-level simulations are conducted using Stim~\cite{gidney2021stim} with a uniform depolarizing noise model. 

\subsection{Selecting Hyperparameters}
To find the smallest viable window size for a given code distance while ensuring $  |W| > |C|  $, we conducted an experiment to analyze optimal commit size as shown in Fig: \ref{fig:commit_size_comp}. Here we keep the buffer size fixed at $  d-1  $ and the commit size $  C  $ is varied over [{1, 2, 3}]. We observed that increasing $  C  $ has negligible impact on LER provided the buffer size is held constant.
A similar analysis was conducted in \cite{Fuhui_Lin_2025} for space-time decoding, where the commit size was held constant and the logical error rate was measured as a function of buffer size across different physical error rates. Their results indicate that increasing the buffer width effectively closes the gap between the target and observed logical error rates.
Combining these insights, and aiming to minimize the window size for the largest possible improvement in reaction time, we fix the commit size to $  C = 1  $ in all subsequent benchmarks.

Avoiding the static overhead, the dynamic hypertuner runs in parallel with the sliding window decoder. For all benchmarks, the initial cutoff is set to $  c_0 = 0.003  $. The hypertuner adjusts the cutoff if $  r_{obs}  $ is not within a target retry range of 20–30\% to ensure that the additional overhead remains below the target decoding cost. 
%\todo{Josh to add QLDPC repo window decoding work description}
\begin{figure}[!ht]
    \centering
    % \begin{subfigure}[b]{0.48\textwidth}
    %     \centering
    %     \caption{BB Codes (baseline $d/2$, target $d$)}
    %     \includegraphics[width=\textwidth]{img/FIG6_bbcode_d2_to_d_ler_per_round_vs_p_inset_rounds30.pdf}
    %     \label{fig:bbcode_d2_to_d}
    % \end{subfigure}
    % \hfill
    % \begin{subfigure}[b]{0.48\textwidth}
    %     \centering
    %     \caption{BB Codes (baseline $d/3$, target $d/2$)}
    %     \includegraphics[width=\textwidth]{img/FIG6_bbcode_d3_to_d2_ler_per_round_vs_p_inset_rounds30.pdf}
    %     \label{fig:bbcode_d3_to_d2}
    % \end{subfigure}
    \includegraphics[width=0.5\textwidth]{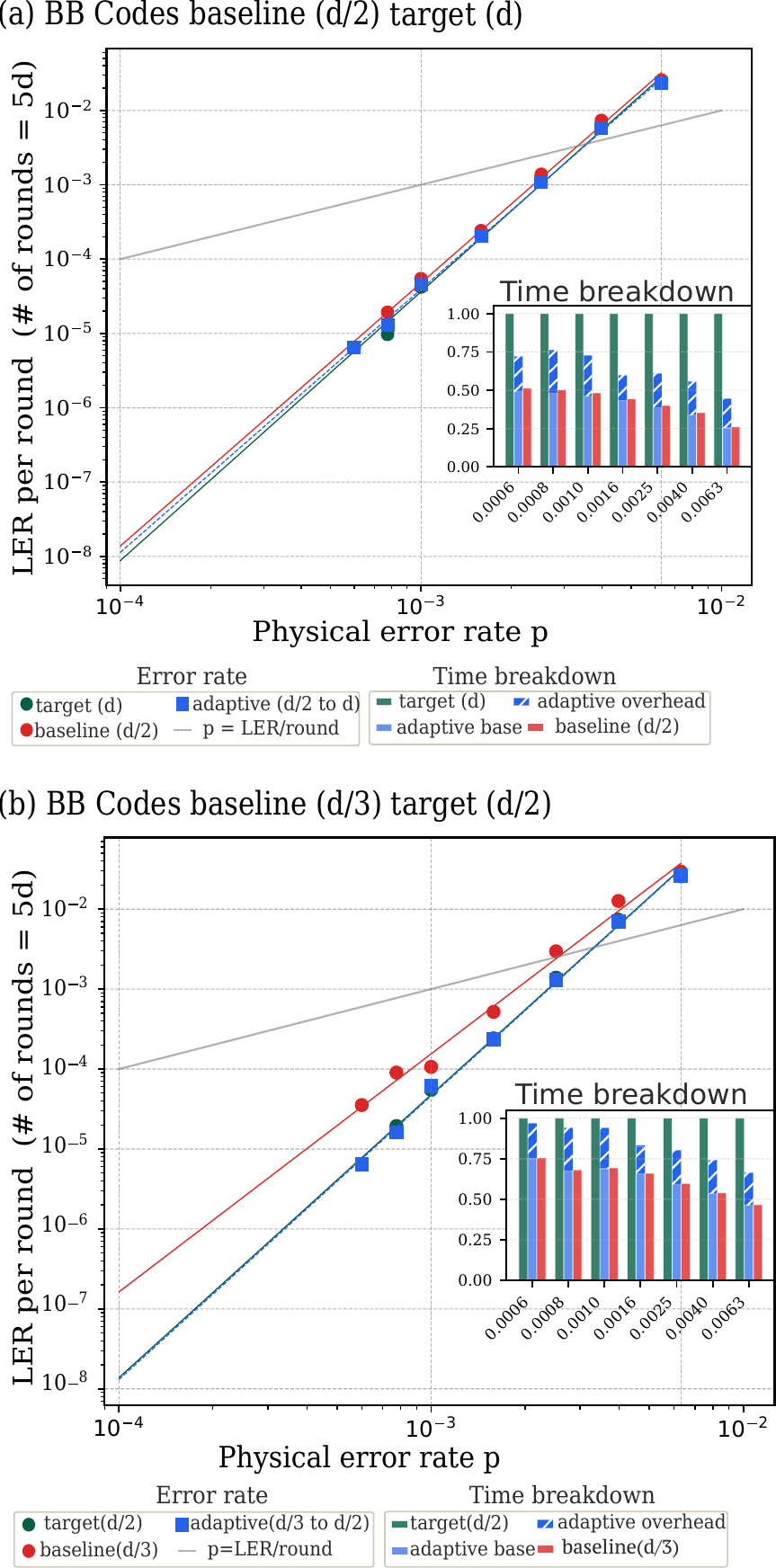}
    \caption{\textbf{Adaptive Window Decoding on Bivariate Bicycle codes } Comparison of decoding in BB Codes $[[72,12,6]]$ distance $d=6$. This figure compares the decoding ler and time for different p. (a) Compares adaptive decoding between target $d$ and baseline $\lfloor d/2 \rfloor$. (Inset) Decoding times normalized relative to the target window size $d$ (maximum time). (B)  Compares adaptive decoding between target $\lfloor d/2 \rfloor$ and baseline $\lfloor d/3 \rfloor$. (Inset) Decoding times normalized relative to the target window size $\lfloor d/2 \rfloor$ (maximum time). }
    \label{fig:decoding_time_bb_codes}
\end{figure}

\begin{figure*}[!htbp]
    \centering
    
    % \begin{subfigure}{0.48\textwidth}
    %     \centering
    %     \caption{NA - LER per round vs p}
    %     \includegraphics[width=\linewidth]{img/FIG7_NA_ler_per_round_vs_p_inset_rounds35.pdf}  % ← replace with your first graph
        
    %     \label{fig:na_ler_vs_p}
    % \end{subfigure}
    % \hfill
    % \begin{subfigure}{0.48\textwidth}
    %     \centering
    %     \caption{NA - LER per round vs rounds}
    %     \includegraphics[width=\linewidth]{img/FIG7_NA_ler_per_round_vs_rounds_errorbars_p0.0015849.pdf}  % ← replace with your second graph
        
    %     \label{fig:na_ler_vs_rounds}
    % \end{subfigure}
    
    % \vskip 1em  % vertical space between rows
    
    % \begin{subfigure}{0.48\textwidth}
    %     \centering
    %     \caption{SI100 - LER per round vs p}
    %     \includegraphics[width=\linewidth]{img/FIG7_SI_ler_per_round_vs_p_time_inset_rounds35.pdf}  % ← replace with your third graph
        
    %     \label{fig:si_ler_vs_p}
    % \end{subfigure}
    % \hfill
    % \begin{subfigure}{0.48\textwidth}
    %     \centering
    %     \caption{SI100 - LER per round vs rounds}
    %     \includegraphics[width=\linewidth]{img/FIG7_SI_ler_per_round_vs_rounds_errorbars_p0.0015849.pdf}  % ← replace with your fourth graph
        
    %     \label{fig:si_ler_vs_rounds}
    % \end{subfigure}
    \includegraphics[width=0.9\textwidth]{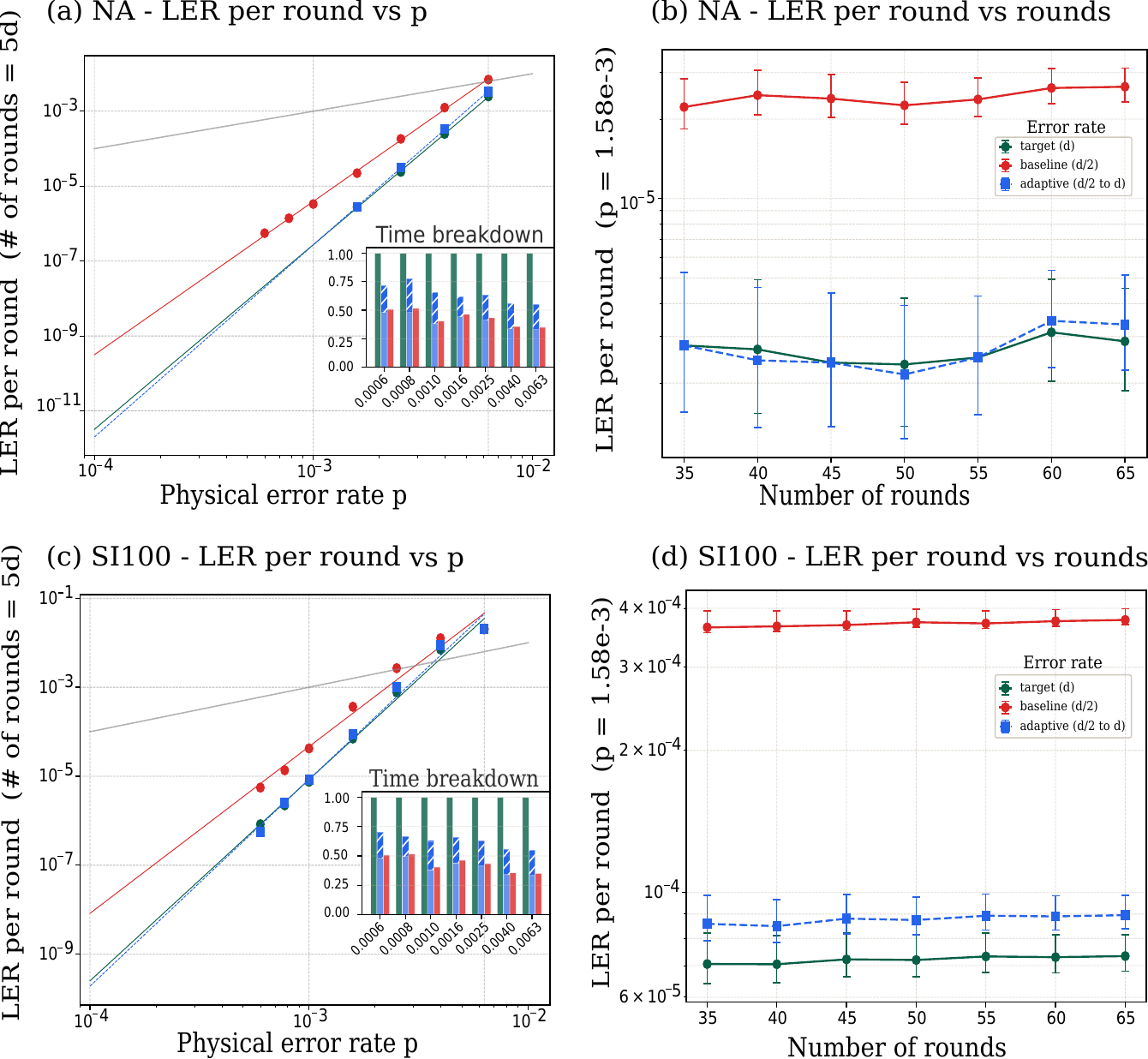}
    \caption{\textbf{Noise sensitivity for the Toric code ($  d=7  $).} The adaptive technique starts with baseline ($W = \lfloor d/2 \rfloor$) and increases window size to target ($W=d$) when decoder has low confidence. (a) Neutral atom inspired noise model. (Inset) Decoding time for all $p$ (b) Logical error rate stability for neutral atom noise model. (c) Superconducting inspired noise model. (Inset) Decoding time for all $p$ (d) logical error rate stability for superconducting noise model.
    For insets, the decoding time is normalised to the maximum decoding time (i.e of target window $  W = d$).}
    \label{fig:noisemodel}
\end{figure*}
\subsection{Performance Results for Toric Codes}
In Fig.~\ref{fig:toric_code_ler_vs_p_45}, we compare the logical error rate (LER) per round and the decoding time for a $d=7$ toric code under depolarizing circuit-level noise, for $35$ syndrome rounds ($\approx 5d$). We evaluate three strategies: the target configuration uses a fixed window size $  W = d  $, the baseline uses a smaller fixed window size $W = \lfloor d/2 \rfloor$, and the adaptive (retry-based) approach starts with the baseline window size $W = \lfloor d/2 \rfloor$ but retries low-confidence windows ($  Q > c  $) using the larger window size $  W = d  $.
For all physical error rates $  p  $, the retry-based method achieves LER nearly identical to that of the target ($  W = d  $). 
For all benchmarks decoding times are normalized to the target window size ($  W = d  $) for each $  p  $. Both the baseline and adaptive schemes exhibit significantly lower decoding time than the target, demonstrating that the improved LER of the retry method comes with substantially reduced overhead compared to using a fixed window of size $  d  $.

\subsection{Performance Results for BB Codes}
We perform the same analysis for Bivariate Bicycle (BB) codes with parameters $  [[72, 12, 6]]  $ under circuit-level depolarizing noise model. As show in Fig~\ref{fig:decoding_time_bb_codes}(a), we evaluate: the target with fixed window size  $W = d$, the baseline with fixed window size $W = \lfloor d/2 \rfloor$, and the adaptive (retry-based) approach that starts with $W = \lfloor d/2 \rfloor$ and retries low-confidence windows using $  W = d  $. Surprisingly, in Fig~\ref{fig:decoding_time_bb_codes}(a) for adaptive ($\lfloor d/2 \rfloor \to d$), the LER gap between the baseline and the target is negligible. We conjecture that this low LER gap between $W = \lfloor d/2 \rfloor$ and $d$ is a result of a greater resilience to measurement noise in BB codes due to some redundancy in checks, allowing BB codes to maintain comparable LER performance even with a reduced window size. 
As a result, the target window size for a distance $d$ BB code may be closer to $\lfloor d/2 \rfloor$ rather than the standard $d$.

To test this, in Fig~\ref{fig:decoding_time_bb_codes}(b), we evaluate the same three schemes for BB codes but for modified window sizes:
the target with fixed window size $W = \lfloor d/2 \rfloor$,
the baseline with fixed window size $W = \lfloor d/3 \rfloor$,
and the adaptive (retry-based) approach, which starts with $W = \lfloor d/3 \rfloor$ and retries low-confidence windows using $W = \lfloor d/2 \rfloor$. Across all physical error rates $  p  $, adaptive window decoding effectively closes the LER gap between the baseline ($W = \lfloor d/3 \rfloor$) and the target ($W = \lfloor d/2 \rfloor$), while maintaining low decoding-time overhead. The reduction in decoding overhead becomes more pronounced as the physical error rate increases. These results demonstrate that the adaptive technique enables across different codes to achieve near-target logical error rates while keeping the overall decoding overhead low. 

\subsection{Noise Model Sensitivity Study}
The adaptive window decoding technique proposed in this work is agnostic to the underlying quantum code and can be easily extended to incorporate future improvements in inner decoders. The threshold confidence $  c_0  $ is a critical parameter in adaptive decoding, as it directly influences overall performance. The dynamic hypertuner updates this threshold to achieve maximum decoding efficiency (keeping the retry rate $r_{obs}$ in check). In this section, we evaluate the effectiveness of the hypertuner within the adaptive window technique across different noise models. Specifically, we investigate its performance under hardware-inspired noise models, since different noise characteristics can substantially impact decoding difficulty. Previously, Fig~\ref{fig:toric_code_ler_vs_p_45} and \ref{fig:decoding_time_bb_codes} discuss adaptive window decoding under depolarizing noise model. In Fig~\ref{fig:noisemodel} we perform a noise sensitivity study for toric codes under neutral atom noise model and superconducting noise model. The noise model definitions are taken from \cite{sahay2026foldtransversalsurfacecodecultivation} for neutral atom and \cite{gidney2024magicstatecultivationgrowing} for superconducting, defined in Table~\ref{tab:noise-models} (where $p$ denotes the base physical error rate).

\begin{table}
\centering
\caption{Hardware inspired noise models}
\begin{tabular}{lcc}
\toprule
\textbf{Operation} & \textbf{NA} & \textbf{SI100} \\
\midrule
Single-qubit Clifford gates & $p/10$ & $p/10$ \\
Two-qubit Clifford gates    & $p$    & $p$ \\
Readout                     & $p$    & $5p$ \\
Reset                       & $p$    & $2p$ \\
Idle                        & $p/10$ & $p/10$ \\
error while waiting for meas/reset & $p/10$ & $2p$ \\
\bottomrule
\label{tab:noise-models}
\end{tabular}
\end{table}

Both noise models differ primarily in readout, reset, and idle errors that occur while waiting for measurement or reset operations. To evaluate the adaptive window technique, we compare its performance under varying physical error rates $  p  $ and numbers of rounds.
Figures~\ref{fig:noisemodel}(a) and~\ref{fig:noisemodel}(b) present results for the neutral-atom-inspired noise model, while Figures~\ref{fig:noisemodel}(c) and~\ref{fig:noisemodel}(d) show results for the superconducting-inspired noise model. In each case, we compare the logical error rate (LER) per round across a range of physical error rates and code distances (number of rounds).
In both noise models, the adaptive window decoding successfully closes the gap in logical error between target and baseline, while maintaining a decoding time substantially lower than that of the conventional $  W = d  $ window decoding scheme. The retry rate remains between 20--30\% (controlled by the dynamic hypertuner), resulting in a decoding time overhead of only 0.4--0.6× relative to the target decoding time across all tested rounds.
For the superconducting-inspired noise model, the LER per round is higher overall, primarily due to elevated reset and readout error rates. Nevertheless, the hypertuning technique proves largely agnostic to the underlying noise model, delivering consistent performance gains across different physical error rates $  p  $ and increasing numbers of rounds.

\begin{figure}[!ht]
    \centering
    \includegraphics[width=0.5\textwidth]{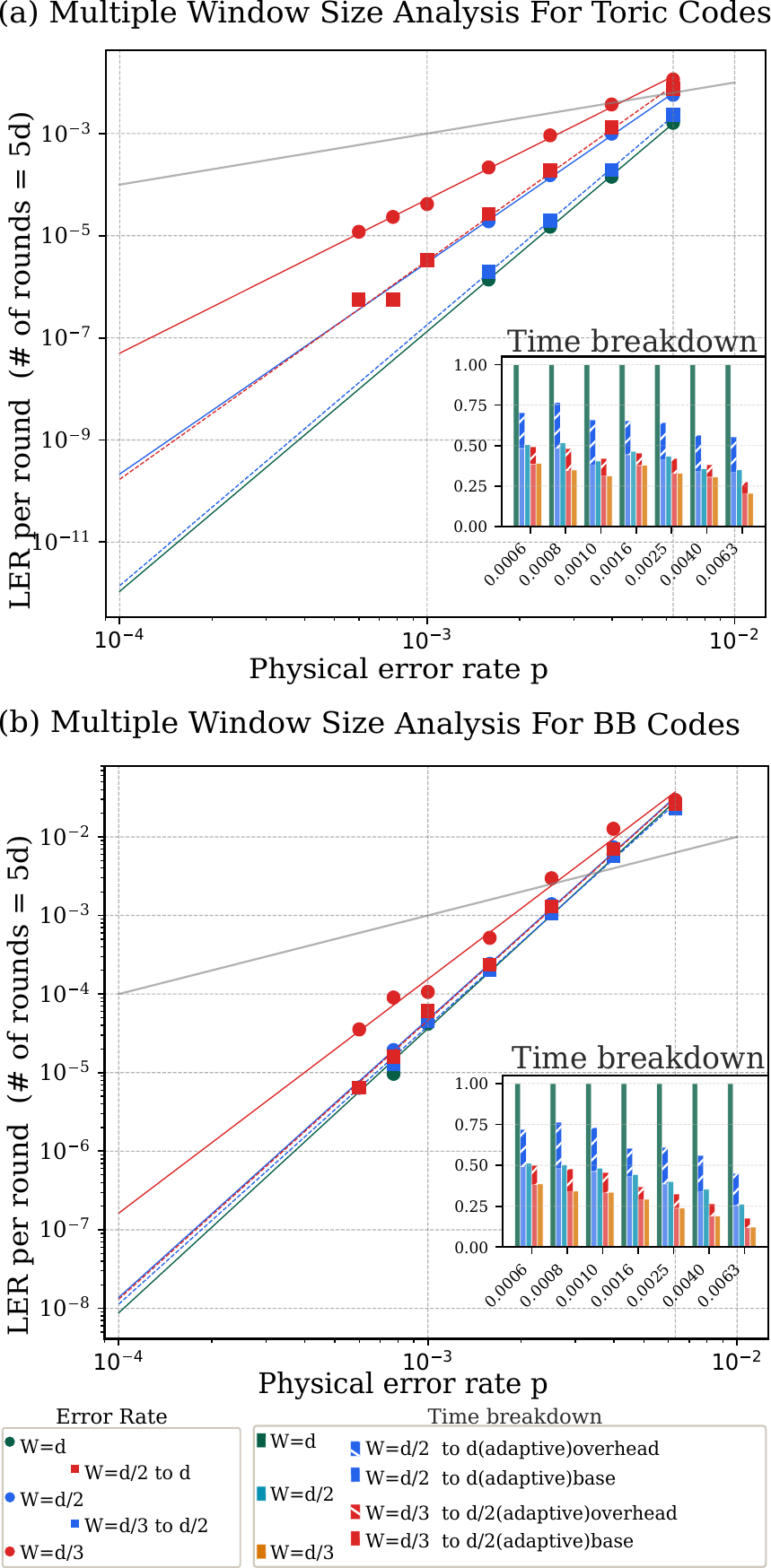}
   
    \caption{\textbf{Performance comparison for different window sizes.}
    This figure compares the performance of fixed target window sizes $w \in \left\{d, \frac{d}{2}, \frac{d}{3}\right\}$ 
    against two dynamic window schemes ($w = \frac{d}{2} \to d$ and $w = \frac{d}{3} \to \frac{d}{2}$) 
    for sliding-window decoding of the (a) toric code and (b) BB codes under circuit-level depolarizing noise 
    at physical error rate $p$.}
    \label{fig:pushing_limits}
\end{figure}
\subsection{Limits for LER toleration}\label{res_sec:C}

In previous sections, we compared the adaptive window technique for different codes and noise models. It is interesting to note that the optimal smaller window size to start can be different for different codes. 
(For example: $\lfloor d/3 \rfloor$ in case of BB codes is more suitable for adaptive window decoding, since $ \lfloor d/2 \rfloor$ and $d$ size window gives similar LER). Furthermore, the precise required logical error rate (LER) can vary from algorithm to algorithm, largely driven by total gate count. Noise-resilient variational algorithms such as VQE and QAOA can tolerate physical error rates of approximately $10^{-2} - 10^{-3}$ per gate without full error correction. Whereas algorithms like, RSA 2048 factoring via Shor's algorithm requires a logical error rate of approximately $10^{-10} - 10^{-12}$ per gate \cite{gidney2025factor2048bitrsa}, reflecting the billions of operations needed to complete the computation reliably. Motivated by this, we also evaluate the performance of our approach for different target logical error rates reached by differing window sizes.. 

In Figure~\ref{fig:pushing_limits}, we analyze the both toric code and BB codes. 
For the Toric code, the adaptive window scheme ($ \lfloor d/3 \rfloor \to \lfloor d/2 \rfloor  $) effectively closes the performance gap between the target window size ($W = \lfloor d/2 \rfloor$) and the more conservative baseline ($W = \lfloor d/3 \rfloor$). The same behavior is observed in the second case, where the adaptive approach ($  \lfloor d/2 \rfloor \to d  $) again bridges the gap to the target performance while maintaining significantly lower decoding time overhead compared to the full $  W = d  $ window. Thus, if a higher LER is deemed tolerable, substantial reductions in reaction time and overall latency can be achieved. In the case of BB codes, the LER gap between the baseline window size ($  \lfloor d/2 \rfloor  $) and the target window size ($  d  $) is already small. However, the adaptive method provides significant improvements when we start with a smaller baseline window of $  \lfloor d/3 \rfloor  $ and increase it to the target window of $  \lfloor d/2 \rfloor  $ for low confidence windows.
This approach allows us to achieve near-optimal LER (close to the target of $  d  $) while significantly reducing the average decoding time for real-time applications. By beginning with a smaller window ($  \lfloor d/3 \rfloor  $) and adaptively increasing it to $  \lfloor d/2 \rfloor  $, we maintain essentially the same LER per round but decode faster overall.

We conclude that in all cases adaptive decoding is able to recover the performance of the target window size while reducing overall decoding time, highlighting the generalizability of our approach. This further complements the inherent trade-offs between logical error rate (LER) and decoding time, allowing the selection of an acceptable LER depending on the requirements of a target program.

\section{Conclusion and Future Work}\label{FAT}\label{sec:conclusion}
In this paper, we address the limitations of fixed window decoding, where a constant large window size imposes unnecessary decoding overhead. Since most error clusters are sparse, such a fixed overhead is often wasteful.
To overcome this, we propose an adaptive window decoding technique. The method uses a cluster-based confidence metric $  Q  $ together with a dynamic hypertuner that adjusts the confidence cutoff in real time based on the observed retry rate.
We show that this approach works robustly across different codes and easily extensible to cluster based decoding algorithms, as it relies solely on decoder soft information. The adaptive technique consistently closes the Logical Error Rate (LER) gap between the baseline (small window) and the target (larger window) while keeping decoding overhead low. These benefits hold across a wide range of physical error rates, noise models, and increasing number of rounds.
For BB codes in particular, we observe that the adaptive method enables further reduction of the window size while still achieving near-target performance.

While our adaptive window technique was demonstrated on sliding window decoding, it is complementary to other window decoding improvements like parallel window decoding~\cite{skoric2023parallel, tan2023scalable} and speculative window decoding~\cite{10.1145/3695053.3731022}. 
Integrating these approaches together is an exciting opportunity for future work which will be particularly important to improve the performance of large-scale, reaction-limited computations~\cite{gidney2025factor2048bitrsa, zhou2025resource, yang2026spacetime}.

% In future work, we intend to extend this adaptive technique to swiper and logical computation. This could help reduce the limitations of reaction-time in decoding for realizing  practical fault tolerance. 

\begin{comment}
In future work, we wish to integrate this approach to logical computation and extending dynamic resizing to speculative decoding frameworks may yield further latency improvements.
\end{comment}

\section*{Acknowledgment}
We thank Siddharth Dangwal for feedback on an earlier version of this work. The authors use AI (Grok, ClaudeAI) to review language and grammar in the paper. All the content for reviewed and edited by the author(s) who take full responsibility of the final work.
This work is funded in part by EPiQC, in part by STAQ under award NSF Phy-Phy-232580; in part by the US Department of Energy Office of Advanced Scientific Computing Research, Accelerated Research for Quantum Computing Program; and in part by the NSF Quantum Leap Challenge Institute for Hybrid Quantum Architectures and Networks (NSF Award 2016136),  in part by the NSF National Virtual Quantum Laboratory program, in part based upon work supported by the U.S. Department of Energy, Office of Science, National Quantum 
Information Science Research Centers, and in part by the Army Research Office under Grant Number W911NF-23-1-0077. The views and conclusions contained in this document are those of the authors and should not be interpreted as representing the official policies, either expressed or implied, of the U.S. Government. The U.S. Government is authorized to reproduce and distribute reprints for Government purposes notwithstanding any copyright notation herein. FTC is the Chief Scientist for Quantum Software at Infleqtion.

%%%%%%%%% -- BIB STYLE AND FILE -- %%%%%%%%
\bstctlcite{bstctl:nodash}
\bibliographystyle{IEEEtranS}
\bibliography{refs}

@IEEEtranBSTCTL{bstctl:nodash,
  CTLdash_repeated_names = "no"
}

@article{hillmann2025localized,
  author = {Hillmann, Timo and Berent, Lucas and Quintavalle, Armanda O. and Eisert, Jens and Wille, Robert and Roffe, Joschka},
  title = {Localized statistics decoding for quantum low-density parity-check codes},
  journal = {Nature Communications},
  volume = {16},
  number = {8214},
  year = {2025},
  doi = {10.1038/s41467-025-63214-7},
  url = {https://www.nature.com/articles/s41467-025-63214-7},
  eprint = {2406.18655},
  archivePrefix = {arXiv}
}

@misc{perlin2023qldpc,
  author = {Perlin, Michael A.},
  title = {{qLDPC}},
  year = {2023},
  publisher = {GitHub},
  journal = {GitHub repository},
  howpublished = {\url{https://github.com/qLDPCOrg/qLDPC}},
}

@software{Roffe_LDPC_Python_tools_2022,
author = {Roffe, Joschka},
title = {{LDPC: Python tools for low density parity check codes}},
url = {https://pypi.org/project/ldpc/},
year = {2022}
}

@misc{lee2026efficientpostselectiongeneralquantum,
      title={Efficient Post-Selection for General Quantum LDPC Codes}, 
      author={Seok-Hyung Lee and Lucas H. English and Stephen D. Bartlett},
      year={2026},
      eprint={2510.05795},
      archivePrefix={arXiv},
      primaryClass={quant-ph},
      url={https://arxiv.org/abs/2510.05795}, 
}

@inproceedings{10.1145/3695053.3731022,
author = {Viszlai, Joshua and Chadwick, Jason D. and Joshi, Sarang and Ravi, Gokul Subramanian and Li, Yanjing and Chong, Frederic T.},
title = {SWIPER: Minimizing Fault-Tolerant Quantum Program Latency via Speculative Window Decoding},
year = {2025},
isbn = {9798400712616},
publisher = {Association for Computing Machinery},
address = {New York, NY, USA},
url = {https://doi.org/10.1145/3695053.3731022},
doi = {10.1145/3695053.3731022},
booktitle = {Proceedings of the 52nd Annual International Symposium on Computer Architecture},
pages = {1386–1401},
numpages = {16},
location = {
},
series = {ISCA '25}
}

@article{skoric2023parallel,
  author = {Skoric, Luka and Browne, Dan E. and Barnes, Kenton M. and Gillespie, Neil I. and Campbell, Earl T.},
  title = {Parallel window decoding enables scalable fault tolerant quantum computation},
  journal = {Nat Commun},
  volume = {14},
  pages = {7040},
  year = {2023},
  doi = {10.1038/s41467-023-42482-1}
}

@article{tan2023scalable,
  author = {Tan, Xinyu and Zhang, Fang and Chao, Rui and Shi, Yaoyun and Chen, Jianxin},
  title = {Scalable Surface-Code Decoders with Parallelization in Time},
  journal = {PRX Quantum},
  volume = {4},
  number = {4},
  pages = {040344},
  year = {2023},
  doi = {10.1103/PRXQuantum.4.040344},
  eprint = {2209.09219},
  archivePrefix = {arXiv}
}

@article{terhal2015quantum,
  author = {Terhal, Barbara M.},
  title = {Quantum error correction for quantum memories},
  journal = {Reviews of Modern Physics},
  volume = {87},
  number = {2},
  pages = {307--346},
  year = {2015},
  doi = {10.1103/RevModPhys.87.307},
  eprint = {1302.3428},
  archivePrefix = {arXiv}
}

@article{dennis2002topological,
  author = {Dennis, Eric and Kitaev, Alexei and Landahl, Andrew and Preskill, John},
  title = {Topological quantum memory},
  journal = {Journal of Mathematical Physics},
  volume = {43},
  number = {9},
  pages = {4452--4505},
  year = {2002},
  doi = {10.1063/1.1499754},
  eprint = {quant-ph/0110143},
  archivePrefix = {arXiv}
}

@article{huang2021between,
  author = {Huang, Shilin and Brown, Kenneth R.},
  title = {Between Shor and Steane: A Unifying Construction for Measuring Error Syndromes},
  journal = {Physical Review Letters},
  volume = {127},
  pages = {090505},
  year = {2021},
  doi = {10.1103/PhysRevLett.127.090505},
  eprint = {2012.15403},
  archivePrefix = {arXiv}
}

@article{iyengar2012windowed,
  author = {Iyengar, Aravind R. and Papaleo, Marco and Siegel, Paul H. and Wolf, Jack K. and Vardy, Alexander and Lentmaier, Michael and Fettweis, Gerhard P.},
  title = {Windowed Decoding of Protograph-Based LDPC Convolutional Codes Over Erasure Channels},
  journal = {IEEE Transactions on Information Theory},
  volume = {58},
  number = {4},
  pages = {2303--2320},
  year = {2012},
  doi = {10.1109/TIT.2011.2177435},
  eprint = {1010.4548},
  archivePrefix = {arXiv}
}

@article{fowler2012surface,
  author = {Fowler, Austin G. and Mariantoni, Matteo and Martinis, John M. and Cleland, Andrew N.},
  title = {Surface codes: Towards practical large-scale quantum computation},
  journal = {Physical Review A},
  volume = {86},
  number = {3},
  pages = {032324},
  year = {2012},
  doi = {10.1103/PhysRevA.86.032324},
  eprint = {1208.0928},
  archivePrefix = {arXiv}
}

@article{delfosse2021almost,
  author = {Delfosse, Nicolas and Nickerson, Naomi H.},
  title = {Almost-linear time decoding algorithm for topological codes},
  journal = {Quantum},
  volume = {5},
  pages = {595},
  year = {2021},
  doi = {10.22331/q-2021-12-02-595},
  eprint = {1709.06218},
  archivePrefix = {arXiv}
}

@article{higgott2022pymatching,
  author = {Higgott, Oscar},
  title = {PyMatching: A Python package for decoding quantum codes with minimum-weight perfect matching},
  journal = {ACM Transactions on Quantum Computing},
  volume = {3},
  number = {3},
  pages = {1--16},
  articleno = {16},
  year = {2022},
  doi = {10.1145/3505637},
  eprint = {2105.13082},
  archivePrefix = {arXiv}
}

@article{bombin2024fault,
  author = {Bombín, Héctor and Pant, Mihir and Roberts, Sam and Seetharam, Karthik I.},
  title = {Fault-Tolerant Postselection for Low-Overhead Magic State Preparation},
  journal = {PRX Quantum},
  volume = {5},
  pages = {010302},
  year = {2024},
  doi = {10.1103/PRXQuantum.5.010302},
  eprint = {2212.00813},
  archivePrefix = {arXiv}
}

@article{gidney2025yoked,
  author = {Gidney, Craig and Newman, Michael and Brooks, Peter and Jones, Cody},
  title = {Yoked surface codes},
  journal = {Nature Communications},
  volume = {16},
  pages = {4498},
  year = {2025},
  doi = {10.1038/s41467-025-59714-1},
  eprint = {2312.04522},
  archivePrefix = {arXiv}
}

@article{bravyi2024high,
  author = {Bravyi, Sergey and Cross, Andrew W. and Gambetta, Jay M. and Maslov, Dmitri and Rall, Patrick and Yoder, Theodore J.},
  title = {High-threshold and low-overhead fault-tolerant quantum memory},
  journal = {Nature},
  volume = {627},
  number = {8005},
  pages = {778--782},
  year = {2024},
  doi = {10.1038/s41586-024-07107-7},
  eprint = {2308.07915},
  archivePrefix = {arXiv}
}

@misc{stein2024architectures,
  author = {Stein, Samuel and Xu, Shifan and Cross, Andrew W. and Yoder, Theodore J. and Javadi-Abhari, Ali and Liu, Chenxu and Liu, Kun and Zhou, Zeyuan and Guinn, Charles and Ding, Yufei and Ding, Yongshan and Li, Ang},
  title = {Architectures for Heterogeneous Quantum Error Correction Codes},
  howpublished = {arXiv preprint arXiv:2411.03202},
  year = {2024},
  eprint = {2411.03202},
  archivePrefix = {arXiv},
  doi = {10.48550/arXiv.2411.03202}
}

@article{Fuhui_Lin_2025,
   title={Spatially parallel decoding for multi-qubit lattice surgery},
   volume={10},
   ISSN={2058-9565},
   url={http://dx.doi.org/10.1088/2058-9565/adc6b6},
   DOI={10.1088/2058-9565/adc6b6},
   number={3},
   journal={Quantum Science and Technology},
   publisher={IOP Publishing},
   author={Fuhui Lin, Sophia and Peterson, Eric C and Sankar, Krishanu and Sivarajah, Prasahnt},
   year={2025},
   month=apr, pages={035007} }

@misc{sahay2026foldtransversalsurfacecodecultivation,
      title={Fold-transversal surface code cultivation}, 
      author={Kaavya Sahay and Pei-Kai Tsai and Kathleen Chang and Qile Su and Thomas B. Smith and Shraddha Singh and Shruti Puri},
      year={2026},
      eprint={2509.05212},
      archivePrefix={arXiv},
      primaryClass={quant-ph},
      url={https://arxiv.org/abs/2509.05212}, 
}

@misc{gidney2024magicstatecultivationgrowing,
      title={Magic state cultivation: growing T states as cheap as CNOT gates}, 
      author={Craig Gidney and Noah Shutty and Cody Jones},
      year={2024},
      eprint={2409.17595},
      archivePrefix={arXiv},
      primaryClass={quant-ph},
      url={https://arxiv.org/abs/2409.17595}, 
}

@misc{gidney2025factor2048bitrsa,
      title={How to factor 2048 bit RSA integers with less than a million noisy qubits}, 
      author={Craig Gidney},
      year={2025},
      eprint={2505.15917},
      archivePrefix={arXiv},
      primaryClass={quant-ph},
      url={https://arxiv.org/abs/2505.15917}, 
}

@article{Higgott_2023,
   title={Improved Decoding of Circuit Noise and Fragile Boundaries of Tailored Surface Codes},
   volume={13},
   ISSN={2160-3308},
   url={http://dx.doi.org/10.1103/PhysRevX.13.031007},
   DOI={10.1103/physrevx.13.031007},
   number={3},
   journal={Physical Review X},
   publisher={American Physical Society (APS)},
   author={Higgott, Oscar and Bohdanowicz, Thomas C. and Kubica, Aleksander and Flammia, Steven T. and Campbell, Earl T.},
   year={2023},
   month=jul }

@misc{fowler2013optimalcomplexitycorrectioncorrelated,
      title={Optimal complexity correction of correlated errors in the surface code}, 
      author={Austin G. Fowler},
      year={2013},
      eprint={1310.0863},
      archivePrefix={arXiv},
      primaryClass={quant-ph},
      url={https://arxiv.org/abs/1310.0863}, 
}

@article{Gidney_2022,
   title={Stability Experiments: The Overlooked Dual of Memory Experiments},
   volume={6},
   ISSN={2521-327X},
   url={http://dx.doi.org/10.22331/q-2022-08-24-786},
   DOI={10.22331/q-2022-08-24-786},
   journal={Quantum},
   publisher={Verein zur Forderung des Open Access Publizierens in den Quantenwissenschaften},
   author={Gidney, Craig},
   year={2022},
   month=aug, pages={786} }

@misc{akahoshi2025runtimereductionlatticesurgery,
      title={Runtime reduction in lattice surgery utilizing time-like soft information}, 
      author={Yutaro Akahoshi and Riki Toshio and Jun Fujisaki and Hirotaka Oshima and Shintaro Sato and Keisuke Fujii},
      year={2025},
      eprint={2510.21149},
      archivePrefix={arXiv},
      primaryClass={quant-ph},
      url={https://arxiv.org/abs/2510.21149}, 
}

@article{panteleev2021degenerate,
  title={Degenerate quantum LDPC codes with good finite length performance},
  author={Panteleev, Pavel and Kalachev, Gleb},
  journal={Quantum},
  volume={5},
  pages={585},
  year={2021},
  publisher={Verein zur F{\"o}rderung des Open Access Publizierens in den Quantenwissenschaften}
}

@article{toshio2025decoder,
  title={Decoder Switching: Breaking the Speed-Accuracy Tradeoff in Real-Time Quantum Error Correction},
  author={Toshio, Riki and Kishi, Kaito and Fujisaki, Jun and Oshima, Hirotaka and Sato, Shintaro and Fujii, Keisuke},
  journal={arXiv preprint arXiv:2510.25222},
  year={2025}
}

@article{gidney2021stim,
  title={Stim: a fast stabilizer circuit simulator},
  author={Gidney, Craig},
  journal={Quantum},
  volume={5},
  pages={497},
  year={2021},
  publisher={Verein zur F{\"o}rderung des Open Access Publizierens in den Quantenwissenschaften}
}

@inproceedings{zhou2025resource,
  title={Resource analysis of low-overhead transversal architectures for reconfigurable atom arrays},
  author={Zhou, Hengyun and Duckering, Casey and Zhao, Chen and Bluvstein, Dolev and Cain, Madelyn and Kubica, Aleksander and Wang, Sheng-Tao and Lukin, Mikhail D},
  booktitle={Proceedings of the 52nd Annual International Symposium on Computer Architecture},
  pages={1432--1448},
  year={2025}
}

@article{yang2026spacetime,
  title={Spacetime-Efficient and Hardware-Compatible Complex Quantum Logic Units in qLDPC Codes},
  author={Yang, Willers and Chadwick, Jason and Teo, Mariesa H and Viszlai, Joshua and Chong, Fred},
  journal={arXiv preprint arXiv:2602.14273},
  year={2026}
}
%%%%%%%%%%%%%%%%%%%%%%%%%%%%%%%%%%%%

\end{document}